\documentclass[final, 5p, times, onecolumn]{elsarticle}

\usepackage{amssymb, amsmath}
\usepackage{textcomp}
\usepackage{subfigure}
\usepackage{braket}
\usepackage{float}
\usepackage{xcolor}
\usepackage{cleveref}
\usepackage{multirow}
\usepackage{icomma}
\usepackage{stackrel}
\usepackage{graphicx}
\usepackage{bm}
\usepackage{url} 

\usepackage{soul}

\usepackage[normalem]{ulem} 

\newcommand{\BG}{\textrm{BG}}

\begin{document}

\title{Dynamics of Cities}

\author{A.~Deppman}
 \address{Instituto de Física -  Universidade de São Paulo, Rua do Matão 1371, São Paulo 05508-090, Brazil
 }
\ead{deppman@usp.br}

\author{R.~L.~Fagundes}
\address{Universidade Federal de Lavras - UFLA, Dept Fis DFI, BR-37200900 Lavras, MG, Brazil }
\ead{fribeiro@ufla.br}

\author{E.~Meg\'{\i}as}
 \address{Departamento de Física Atómica, Molecular y Nuclear and Instituto Carlos I de Física Teórica y Computacional, \\
 Universidad de Granada, Avenida de Fuente Nueva s/n, 18071 Granada, Spain
 }
\ead{emegias@ugr.es}

\author{R.~Pasechnik}
\address{Department of Physics, Lund University, S\"olvegatan 14A,
Lund SE-22362, Sweden}
\ead{Roman.Pasechnik@fysik.lu.se} 

\author{F.~L.~Ribeiro}
\address{Universidade Federal de Lavras - UFLA, Dept Fis DFI, BR-37200900 Lavras, MG, Brazil }
\ead{fribeiro@ufla.br}

\author{C.~Tsallis}
 \address{Centro Brasileiro de Pesquisas Fisicas and National Institute of Science and Technology of Complex Systems, \\Rua Xavier Sigaud 150, Rio de Janeiro-RJ 22290-180, Brazil \\ Santa Fe Institute, 1399 Hyde Park Road, Santa Fe, 87501 NM, USA\\
 Complexity Science Hub Vienna - Josefst\"adter Strasse 39, 1080 Vienna, Austria}
\ead{tsallis@cbpf.br}

\begin{abstract}
This study investigates city dynamics employing a nonextensive diffusion equation suited for addressing diffusion within a fractal medium, where the nonadditive parameter, $q$, plays a relevant role. The findings demonstrate the efficacy of this approach in determining the relation between the fractal dimension of the city, the allometric exponent and $q$, and elucidating the stationary phase of urban evolution. The dynamic methodology facilitates the correlation of the fractal dimension with both the entropic index and the urban scaling exponent identified in data analyses. The results reveal that the scaling behaviour observed in cities aligns with the fractal dimension measured through independent methods. Moreover, the interpretation of these findings underscores the intimate connection between the fractal dimension and social interactions within the urban context. This research contributes to a deeper comprehension of the intricate interplay between human behaviour, urban dynamics, and the underlying fractal nature of cities.
\end{abstract}

\maketitle

\section{Introduction}

\subsection{The city's fractal space}

In recent decades, our understanding of urban population dynamics has rapidly advanced. Exploiting precise data made available by social media platforms and mobile devices, contemporary urban issues can now be scrutinized through more rigorous scientific methodologies. The development of new analytical methods for studying complex systems, previously inaccessible for the study of urban systems, has facilitated the quantitative and qualitative comprehension of various facets of urban life.

This advancement has brought forth intriguing aspects of urban organization. Contrary to prior assumptions, cities exhibit some adherence to universal laws, independent of cultural, ethnic, or socioeconomic nuances of their specific regions~\cite{Bettencourt2007, Batty2013,West2018-xi}. Numerous aspects of urban existence follow simple power laws with population, manifesting either superlinear or sublinear tendencies, a phenomenon known as \textit{urban scaling}. Socioeconomics displays superlinear trends, while infrastructure-related aspects demonstrate sublinear behavior, allowing for enhanced efficiency with population growth. The associated exponent remains remarkably consistent across diverse cities, regardless of their unique attributes~\cite{Rybski2019}. Despite robust evidence supporting the power-law dynamics in urban parameters concerning population, our understanding of the underlying mechanisms driving urban growth remains limited~\cite{Bettencourt2020,Cao2023}.

The various models attempting to elucidate urban scaling were summarized in Ref.~\cite{Ribeiro2023}, emphasizing their reliance on human interaction and the availability of infrastructure. These models diverge in their approaches to deriving scaling exponents, which are invariably obtained by considering allometric relations between socioeconomic output and infrastructure cost~\cite{Molinero2021}. The observed universal behaviour results from constraints related to the stability of the urban area~\cite{Bettencourt2013}. A range of mechanisms have been proposed, such as cross-sectional interaction \cite{Bettencourt2013}, gravitational model~\cite{Ribeiro2017} and other preferential attachment approaches~\cite{Yang2019} (see also \cite{CamachoVidales2023}). It is worth mentioning that preferential attachment is a known mechanism to generate networks with power-law behaviour~\cite{Barabasi1999}. Fractal dimensions, either linked to social behaviour or to infrastructure distribution, are commonly employed to obtain the appropriate range for the power-law exponent~\cite{Molinero2021,Bettencourt2013,Yakubo2011,Arbesman2009}. Most of the models describe static properties of the cities, but there is evidence that dynamical effects are relevant~\cite{Pumain2006,Rybski2016}. The present work addresses the dynamical evolution of the cities by considering how they are modified as the population distribution changes with time. One of the most important relations is the so-called \textit{fundamental allometry}, relating the city area $A$ to the population size $N$ by
\begin{equation}
    A \sim N^{\beta}\,, \label{eq:allometry}
\end{equation}
with $\beta$ being the scaling exponent.

\subsection{Nonlinear dynamics in fractal space}

The dynamics approach to urban life allows for a more comprehensive understanding of the organic organization of individuals and infrastructure. This approach encompasses the city's geometry and socioeconomic interactions within a single theoretical framework. Some attempts to develop a dynamical theory of the cities were associated with Levy-flights, but they do not reproduce some important aspects observed in cities around the world~\cite{Xu2021-ah}. The results obtained in this work will show that the non-additive entropy and the associated non-extensive thermodynamics~\cite{Tsallis} offer a better framework for describing urban life.

Most of the urban scaling studies select cities above some minimum population size, while some works exhibit that small cities diverge from the expected pattern~\cite{Meirelles2018}.
A model for information diffusion in fractal networks~\cite{DeppmanPLOSOne2021} shows that the scaling law may fail for small groups. The behaviour of information spreading with the population size follows the $q$-exponential function instead of the power-law function. The 
$q$-exponential function is 
given by~\cite{Tsallis}
\begin{equation}
\begin{cases}
    e_q(x) = \left[1-(q-1)x\right]^{\frac{-1}{q-1}} \,,  \qquad x < 1/(q-1)\\
    e_q(x) = 0 \,, \qquad\qquad\qquad\qquad\; x \ge 1/(q-1)
\end{cases}    \,,
\end{equation}
where $q \in \mathbb{R}$ is the entropic index; this work focuses on the case $q>1$. Both the power-law and the $q$-exponential functions exhibit similar behaviour for large population sizes but differ for small ones. The $q$-exponential distribution is typical of nonextensive statistics~\cite{Tsallis}, which generalizes Boltzmann's statistics by allowing a non-additive entropy. The generalized statistics has found numerous applications in many realms of knowledge~\cite{TsallisBook,Tsallis2023}. The relationship between fractals and non-extensive statistics has been explored in several works~\cite{Alemany1994,Lyra1998,Deppman-Physics-2021,golmankhaneh2021tsallis}.

Assuming the validity of the fractal model for urban landscapes, a comprehensive exploration of this structure may uncover key aspects for predicting cities' dynamic behaviour. This approach has the potential to yield vital information about cities' temporal evolution, thus presenting a valuable opportunity to devise more effective strategies for urban development. 

The relevant time scale for attaining the stationary regime assumed here is one in which the characteristic exponent of the distributions can be considered constant. Evidence indicates that for time scales spanning centuries, the exponent may change, but the present work is focused on the dynamics of cities in a shorter term. This approach encompasses processes such as the immigration of representative fractions of the population of the city or the relocation of the population due to natural disasters. These processes can disrupt the natural population density of the city, triggering a subsequent change through a dynamic process that must incorporate the inherent fractal aspects of the city's organic evolution.

The dynamics of systems in fractal spaces are rather different from the usual dynamic evolution. The \textit{Fokker-Planck Equation} (FPE) is the law one usually has in mind when addressing the evolution of a complex system, and if $f(\mathbf{r},t)$ is the probability distribution, which depends on the position $\mathbf{r}$ and time $t$, then the FPE is given by
\begin{equation}
 \frac{\partial f}{\partial t}(\mathbf{r},t)=  \frac{\partial }{\partial x_i} \left[ -\gamma_i(\mathbf{r}) f(\mathbf{r},t) + B \frac{\partial }{\partial x_i} f(\mathbf{r},t) \right] \,, \label{eq:FPE}
\end{equation}
where summation over index $i$ is understood. In the FPE, the parameters $\gamma_i(\mathbf{r})$ and $B$ are transport coefficients and characterize the drift of the system in the medium and the diffusive process through the medium, respectively\footnote{In general, the transport coefficients can assume a tensor form. The assumption that they are scalar, in particular $B_{ij} = B \, \delta_{ij}$, is appropriate for the application we intend here concerning an isotropic system. Tensor coefficients can be relevant for situations where the city growth is constrained by geographical features.}.

\section{Dynamics in non-homogeneous media}

Non-homogeneous media may give rise to a modified process that a non-linear Fokker-Planck Equation can describe. The one-dimensional case where $\bm{\gamma}(\mathbf{r}) = \bm{\gamma}_1 -\gamma_2 \mathbf{r}$ is of particular interest and has been addressed in Ref.~\cite{TsallisBukman} through a comprehensive study of anomalous diffusion. The parameter $\bm{\gamma}_1$ indicates a constant repulsive force, while the parameter $\gamma_2$ is associated with an attractive harmonic potential that represents the overall tendency of the population to live near some basic facilities offered by urban centres. The harmonic potential yields an area for the city that increases linearly with the population size if fractal effects are absent.

The Plastino-Plastino Equation (PPE)~\cite{PLASTINO1995347} is a quite general equation for non-linear dynamics and is particularly useful when a fractal medium is present. This equation is given by
\begin{equation}
  \frac{\partial f}{\partial t}(\mathbf{r},t)=  \frac{\partial }{\partial x_i} \left[ -\gamma_i(\mathbf{r}) f(\mathbf{r},t) + B \frac{\partial }{\partial x_i} f(\mathbf{r},t)^{2-q} \right] \,, \label{eq:PPE}
\end{equation}
where $q$ is the entropic index of nonextensive statistics~\cite{TsallisBook}. The PPE was proposed in the context of nonextensive statistics~\cite{Tsallis,TsallisBook}, therefore it is the appropriate framework to describe the evolution of a wide class of complex systems, and its solutions present $q$-exponential forms of distribution or related ones~\cite{Plastino2024}. The PPE has been recently derived from a generalized form of the Boltzmann Equation for systems with non-local correlations~\cite{Deppman2023}. The dynamics of systems in fractal space can be related to the PPE and in this case, $q$ can be determined by the fractal dimension of the system~\cite{Deppman2023,MGD2024}, highlighting the close connections between fractals and nonextensive statistics.

In an alternative approach, when a system evolves in a fractal medium, the FPE may be modified by substituting the standard derivative operators with \textit{fractal derivatives}. The result is the Fractal Fokker-Planck Equation (FFPE) given by~\cite{Golmankhaneh2022-vn}
\begin{equation}
 D^{\alpha^\prime}_{t_o} f(\mathbf{r},t)=  D^{\alpha^{\prime \prime}}_{x_{i,o}} \left[ -\gamma_i(\mathbf{r}) f(\mathbf{r},t) + B D^{\alpha^{\prime \prime}}_{x_{i,o}} f(\mathbf{r},t) \right] \,, \label{eq:FFPE}
\end{equation} 
where $0<\alpha^\prime \le 1$ and
$0<\alpha^{\prime \prime} \le 1$ represent the fractal dimension of the time and coordinate spaces, respectively. 
It is important to highlight the distinction between \textit{fractal derivatives} and \textit{fractional derivatives} at this point. The former is associated with Haussdorff geometry~\cite{FD1}, while the latter class of derivative operators is derived from algebraic considerations of derivative operators. The connection between these two derivative classes can be established through the continuous approximation of the fractal derivative~\cite{DMP2023}.

One of the possible continuous approximations is associated with $q$-deformed calculus~\cite{Borges-qCalculus}. This approximation transforms the fractal version of the Fokker-Planck Equation into the Plastino-Plastino Equation given by Eq.~(\ref{eq:PPE}) \cite{MGD2024}. Exploiting the properties of the $q$-calculus derivative, Ref.~\cite{MGD2024} demonstrated that the most significant geometrical aspect influencing the dynamic process in the fractal space is the fractal dimension gap, denoted as $\delta d_f \equiv d-d_f$, where $d$ is the smallest integer dimension of Euclidean space that embeds the fractal space with dimension $d_f$.

A consequence of this finding is that the effects of time and coordinate fractal spaces depend on the joint space $(t,\mathbf{r})$. The dimension of the joint space is the sum of the time and coordinate spaces, implying that the fractal dimensions $\alpha^{\prime}$ and $\alpha^{\prime \prime}$ in Eq.~(\ref{eq:FFPE}) can be substituted by $\alpha=\alpha^{\prime}+\alpha^{\prime \prime}$.

Several works have explored the fractional version of the PPE~\cite{Bologna2000,Lenzi2009}. The present work addresses the connections between the fractal equation and the PPE by adopting the continuous approximation associated with the $q$-deformed calculus. The standard derivative formulas can be recovered by using such continuous approximations, resulting in the PPE equation. In the connection between these two equations, the important quantity $\xi$ was derived as
\begin{equation}
\xi = 2 - d(q-1) = 2 - \delta d_f  \in (0,2]\,, \label{eq:chi-fractalgap}
\end{equation}
establishing the link between fractal geometry, the dynamics in fractal space and the nonextensive dynamics. The quantity $\xi$ is defined to be the fractal dimension $\alpha$ of the image space of a function. For a distribution that is positively defined, $\alpha=\xi/2$.

Eq.~(\ref{eq:chi-fractalgap}) results from considerations on the foundations of fractal geometry, fractal and $q$-deformed calculus~\cite{MGD2024} and nonlinear dynamics. It allows for the calculation of the parameter $q$ for any fractal space, since 
\begin{equation}
    q  = 2- d_f/d \,. \label{eq:qdeltaf}
\end{equation}
The PPE allows for addressing the dynamic aspect of the cities.

\section{Dynamics of cities}

The present work assumes that the city growth is isotropic, that is, the infrastructure expands with the same probability in any direction concerning the city centre $\mathbf{r}_c(t)$ so that the probability distribution will be a function of $|\mathbf{r} - \mathbf{r}_c(t)|$. This assumption allows for a simplified equation (see Eq.~(\ref{eq:isocityPPE}), below). The solution of the PPE~(\ref{eq:PPE}) corresponding to a Dirac $\delta$-distribution at $t=0$  is given by 
\begin{equation}
 f(\mathbf{r},t)= \frac{1}{N(t)} e_q\left[-\frac{(\mathbf{r}-\mathbf{r}_c(t))^2}{2\sigma(t)^2}\right]\,, \label{eq:qgauss}
\end{equation}
where
\begin{equation}
 \begin{cases}
   \mathbf{r}_{c}(t)= \frac{\bm{\gamma}_1}{\gamma_2}+\left(\mathbf{r}_{o}-\frac{\bm{\gamma}_1}{\gamma_2}\right) \exp[-\gamma_2 t]  \, \\
  \\
\sigma(t) = \sigma_\infty \left(  1  -  \exp\left[ -\xi \gamma_2 t \right] \right)^{\frac{1}\xi} 
 \end{cases}\,,
 \label{soltutionparameters_d}
\end{equation}
with 
\begin{equation}
\sigma_\infty \equiv \ell_q \, \kappa^{\frac{1}{\xi}} \,, \qquad
    \kappa \equiv (2-q) (2\pi \chi_q)^{\frac{d}{2}(q-1)} \frac{B}{\gamma_2 \ell_q^2}  \,. \label{eq:sigma-kappa}
\end{equation}
 The parameter $\ell_q$ is 
{\it the characteristic linear size of the fractal space}, and
\begin{equation}
\chi_q \equiv \frac{1}{q-1} \left( \frac{\Gamma\left( \frac{1}{q-1} - \frac{d}{2}\right) }{\Gamma\left( \frac{1}{q-1} \right)} \right)^{\frac{2}{d}}  \,, \qquad q > 1 \,. \label{eq:chiq_d}
\end{equation}
The dimension $d$ remains arbitrary because the same method used here can be applied to other processes with different dimensions. The solution in Eq.~(\ref{eq:qgauss}) is a $q$-Gaussian, with a peak at $\mathbf r = \mathbf r_c(t)$ and width $\sigma(t)$. For $t\rightarrow \infty$, the city reaches a new stationary regime after being disturbed by some event of the kind mentioned before. Observe that the distribution in Eq.~(\ref{eq:qgauss}) is dimensionless, as required for the correct usage of the PPE because all parameters are written in scaling invariant form, such as $\sigma/\ell_q$. Below, the dimensional distribution is recovered.

To develop a dynamical model of cities, it is important to observe that cities usually start and evolve around a fixed geometric centre, which remains, to a good approximation, the centre of mass of the urban area during its evolution. In the following, it will be assumed that the city centre is at $\mathbf r_c(t) = \mathbf 0 $. This condition is obtained by considering $\mathbf{r}_o = \mathbf{0} = \bm{\gamma}_1$.  This means that one can integrate in the angular coordinates, and the distribution will depend only on $r$ and $t$, i.e., $f(r,t)$. The PPE for the isotropic city becomes
\begin{eqnarray}
 \frac{\partial}{\partial t} f(r,t) &=&  \gamma_2 \left( d + r \frac{\partial }{\partial r} \right) f(r,t) \nonumber \\
 &&+ \frac{B}{r^{d-1}} \frac{\partial}{\partial r} \left( r^{d-1} \frac{\partial}{\partial r} f(r,t)^{2-q} \right) \,, \label{eq:isocityPPE}
\end{eqnarray}
and its solution is given by Eq.~(\ref{eq:qgauss}) with $\mathbf r_c(t) = \mathbf 0$. Notice that the function $f(r,t)$ is the profile of the population density at the time $t$, measured from the city centre, $r=0$. 

While the distribution $f(\mathbf{r},t)$ is dimensionless, the rescaled distribution
\begin{equation}
\bar f(\mathbf{r},t) \equiv \frac{N(t)}{\ell_q^d} f(\mathbf{r},t)
\end{equation}
has dimensions of $\textrm{[Length]}^{-d}$ and it is interpreted as the population density. The integral of the density over all space gives the population size, i.e.,
\begin{equation}
    N(t) =  \int d^d r \bar f(\mathbf r, t)  \,, \label{eq:popsize}
\end{equation}
resulting in the total population at time $t$,
\begin{equation}
N(t) = \left( \sqrt{2\pi \chi_q} \frac{\sigma(t)}{\ell_q} \right)^d \,. \label{eq:Nsigmat} 
\end{equation}

Observe that the stationary distribution has a finite width for $t \rightarrow \infty$, and the population at this stage will be
\begin{equation}
    N_{\infty}(q)= (2\pi \chi_q)^{d/2} \kappa^{d/\xi}\,,
\end{equation}
where the dependence in the parameter $q$ is evidenced. The finite width at the asymptotic stationary regime is a consequence of the harmonic potential~\cite{TsallisBukman}.
The parameters $B$ and $\gamma_2$ do not depend on $q$, so the time dependence of the width is also independent of $q$, as can be observed in Eq.~(\ref{soltutionparameters_d}).

A paradigmatic case is $q=1$, when the PPE and FFPE reduce to the FPE, the solution becomes a Gaussian, and the whole process is governed by Boltzmann statistics in a Euclidean space, as indicated in Eq.~(\ref{eq:chi-fractalgap}). For this case, the stationary population will be
\begin{equation}
    N_{\BG} \equiv N_\infty(q=1)=(2\pi)^{d/2}~\kappa_{\BG}^{d/2}\,.
\end{equation}

\begin{figure*}[t]
\centering  
\includegraphics[width=0.23\textwidth]{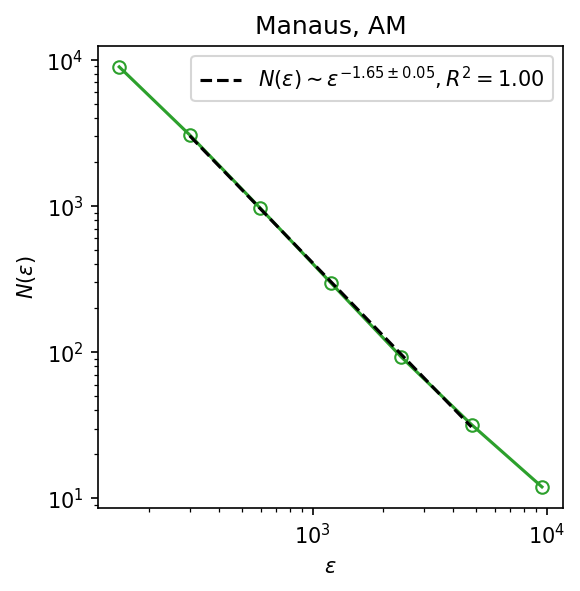} 
 \centering
\includegraphics[width=0.23\textwidth]{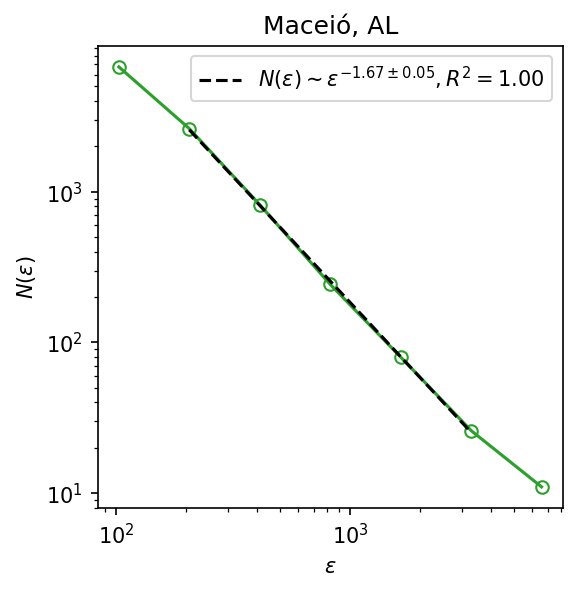} 
\includegraphics[width=0.23\textwidth]{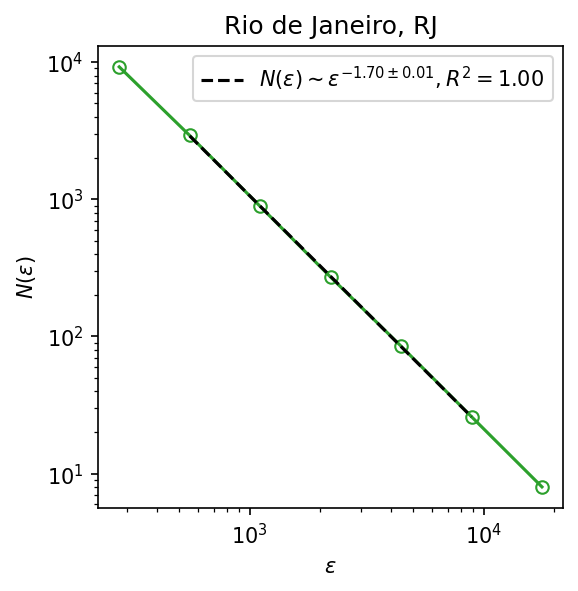} 
 \centering
\includegraphics[width=0.23\textwidth]{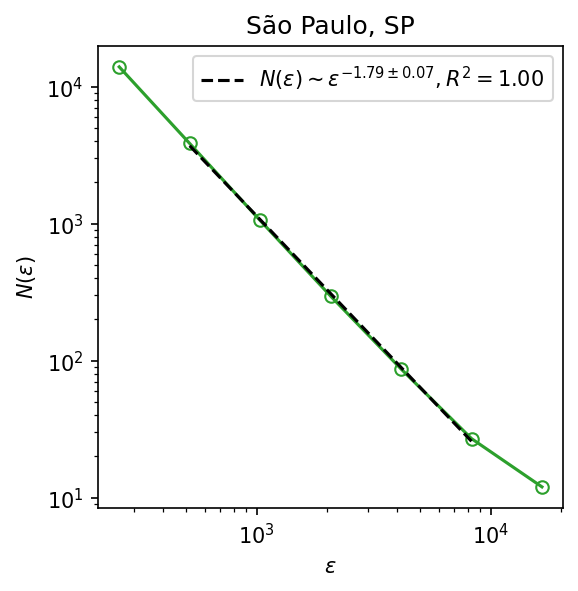} 
\centering
\includegraphics[width=0.23\textwidth]{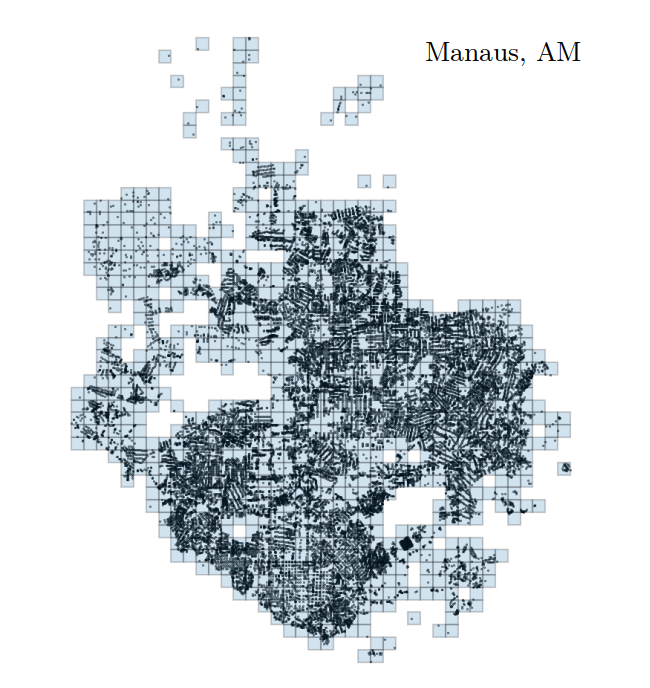}
\centering
\includegraphics[width=0.23\textwidth]{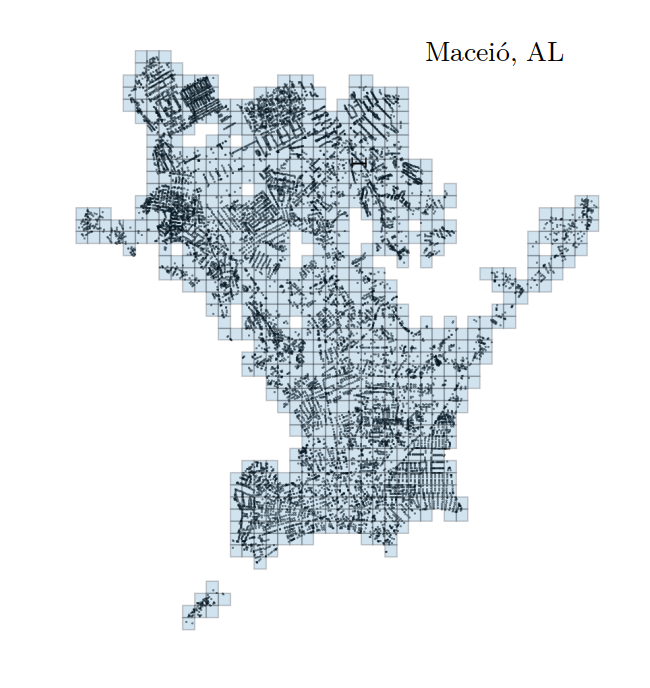}
\centering
\includegraphics[width=0.23\textwidth]{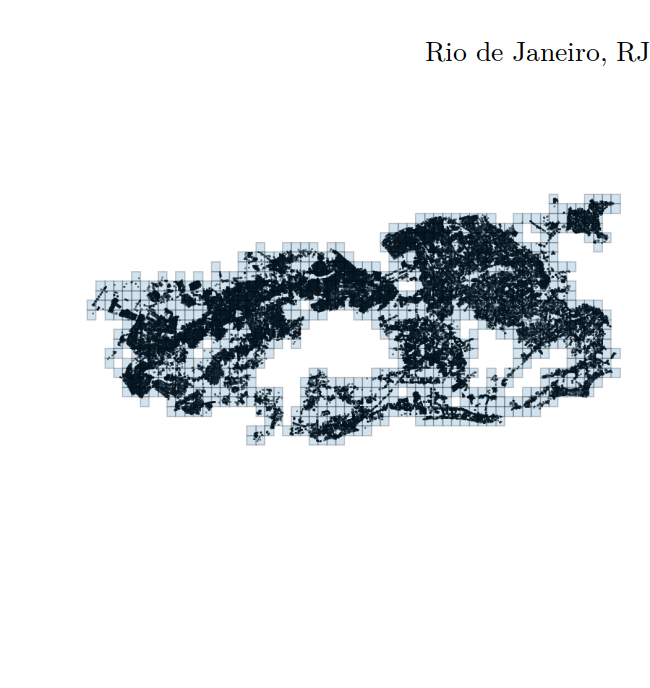}
\centering
\includegraphics[width=0.23\textwidth]{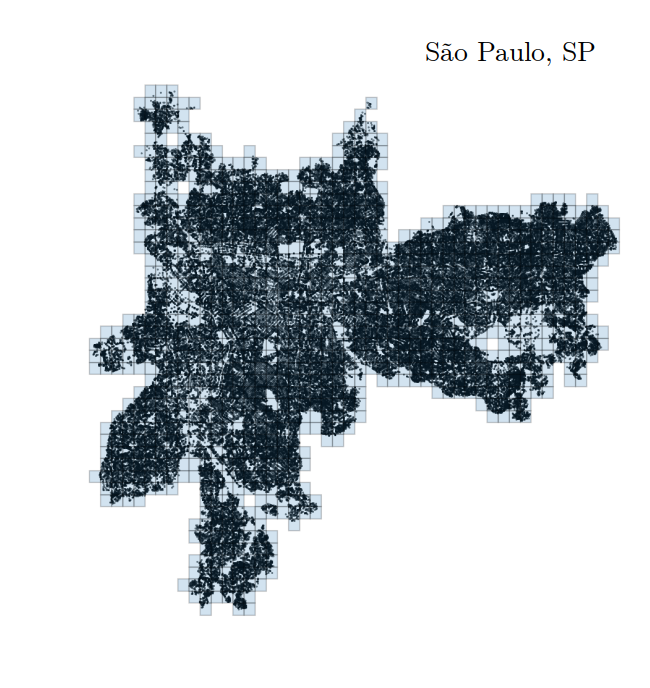}

 \caption{
 Typical plots for the determination of the fractal dimension $d_f$. Here, the representative cases are for the Brazilian cities: Manaus ($d_f=1.65$, nodes $= 28652$, $N_{\ell_{i=6}} = 966$, $\ell_{i=6}=597$ m). Maceió ($d_f=1.67$, nodes $= 15248$, $N_{\ell_{i=6}} = 821$, $\ell_{i=6}=411$ m), Rio de Janeiro ($d_f=1.70$, nodes $= 70329$, $N_{\ell_{i=6}} = 899$, $\ell_{i=6}=1107$ m) and São Paulo ($d_f=1.79$, nodes $= 113272$, $N_{\ell_{i=6}} = 1054$, $\ell_{i=6}=1036$ m). The dashed line indicates the region used for the linear fitting, and the green line is for guiding-eye purposes.}
\label{fig:plotDf}
\end{figure*}

The crucial point is to find the conditions to have the same population in a $q$-Gaussian distribution for any value of $q$, so one can study the effects of the different values for the fractal dimensions and the entropic index on the population density. From Eq.~(\ref{eq:sigma-kappa}), it results that to keep a constant population, $\ell_q$ must vary with $q$ in such a way that
\begin{equation}
    \sqrt{2\pi \chi_q} \frac{\sigma(q)}{\ell_q}= \sqrt{2\pi \chi_q} \, ~\kappa^{1/\xi} \label{eq:sigmaellkappa}
\end{equation}
remains independent of $q$. By using Eq.~(\ref{eq:sigma-kappa}) it follows that
\begin{equation}
    \sqrt{2\pi \chi_q} \frac{\sigma(q)}{\ell_q}= \sqrt{2\pi} \left( \frac{u_q \ell_o}{\ell_q} \right)^{2/\xi}\,, \label{eq:Intermediate1}
\end{equation}
where
\begin{equation}
u_q =  \left[ ~(2-q) (2\pi)^{ \frac{2 - \xi}{2}}\chi_q \right]^{1/2} \,, \label{eq:uq}
\end{equation}
with $\ell_o=\sqrt{B/\gamma_2}$ being a typical length associated with the dynamic properties of the system, as discussed below.

The population size $N(t)$ is independent of $q$ if the right-hand side of Eq.~(\ref{eq:Intermediate1}) is $q$-independent. In this case, the equality
\begin{equation}
    \sqrt{2\pi} \left( \frac{u_q \ell_o}{\ell_q} \right)^{2/\xi}=\sqrt{2\pi} \frac{\sigma_{\BG}}{\ell_{\BG}}\,,
\end{equation}
with $\sigma_{\BG}=\sigma_{q=1}$ and $\ell_{\BG}=\ell_{q=1}$, must hold, implying that
\begin{equation}
    \frac{u_q \ell_o}{\ell_q}=\left( \frac{\sigma_{\BG}}{\ell_{\BG}} \right)^{\xi/2} \,.
\end{equation}
Observe that, for $q=1$, it results $u_q=1$ therefore $\sigma_{\BG}=\ell_o$. Thus, the length $\ell_o$ represents the distribution width for the solution of the Fokker-Planck Equation associated with the diffusive process governed by the Boltzmann Statistics.

Establishing the scaling relation
\begin{equation}
        \ell_q = u_q \ell_o^{1-\xi/2} \ell_{\BG}^{\xi/2} \,, \label{eq:scaling-l}
\end{equation}
ensures that $\kappa^{2/\xi}$ in Eq.~(\ref{eq:sigmaellkappa}) is independent of $q$. Thereby, $\sigma_q$ scales with $q$ in the same way as $\ell_q$, so $\sigma(q) \propto \ell_{\BG}^{\xi/2}$. Since, for the same reason, $\sigma_{\BG} \propto \ell_{\BG}$, it follows that
\begin{equation}
 \sigma(q) \propto \sigma_{\BG}^{\xi/2} \,, \label{eq:scaling-sigmaS}
\end{equation}
showing the scaling behaviour of the distribution width. The population density
\begin{equation}
    \rho_q(t)=N(t)/\ell_q^d \,,
\end{equation}
increases as the fractal dimension decreases. The area of the city is $A_q=\ell_q^d$ and scales in the same way as the linear length, i.e., $A_q=A_{\BG}^{\xi/2}$. But $A_{\BG} \propto N$, therefore
\begin{equation}
    A_q \propto N^{\xi/2}\,.
\end{equation}
By comparing the expression above with the fundamental allometry, it results that the infrastructure scaling exponent is $\beta = \xi/2$.

The dynamic theory in a fractal space yields consistent beha\-viour in linear scales, as can be observed by comparing Eqs.~(\ref{eq:scaling-l}) and (\ref{eq:scaling-sigmaS}). Both $\ell_q$ and $\sigma(q)$ scale according to the same power-law. Note that this scaling is valid for any time $t \ne 0$, indicating that it is a feature of the fractal space that induces non-local correlations into the evolution of the system, thereby leading to non-additive statistics. It is worth understanding the role of the parameter $\ell_q$ in the dynamic process. It was named the characteristic linear size of the fractal space because it represents the scale of the fractal space with dimensional gap $\delta d_f$. It imprints the same power-law trend to the distribution width, therefore it is through  $\ell_q$ that the system inherits the scaling properties observed in the dynamical parameter $\sigma(t)$. 

Using Eqs~(\ref{eq:chi-fractalgap}) and~(\ref{eq:scaling-l}), it results that the characteristic linear size of the system varies with $q$ and the fractal dimension gap as a power-law with exponent
\begin{equation}
    \beta=\frac{\xi}{2}=1-\frac{d}{2}(q-1)=1-\frac{d- d_f}{2}\, , \label{eq:scalingexp}
\end{equation}
where $\beta$ is the scaling exponent of the linear size of the dynamic distribution, $q$ is the entropic index of the Tsallis Statistics, and $d_f$ being the fractal dimension. By using $d = 2$, it results from Eq.~(\ref{eq:scalingexp}) that $\beta = d_f/2$. As we will discuss below, a reasonable value for the city's fractal dimension is $d_f = 1.7$, resulting in $\beta = 0.85$. We will see that this value for the scaling exponent is also in good agreement with empirical findings.

The results obtained above show that starting from the dynamics on a fractal space, in a process that evolves diffusively under harmonic attractive forces, the sub-linear behaviour of the urban infrastructure is obtained. This is in contrast with the observed for the FPE solutions, corresponding to the Boltzmannian systems, which present a linear relation between urban area and population size. This behaviour is recovered when the $d_f \rightarrow 2$.
Thus, the fundamental allometric relation for cities is reproduced in an approach that considers only the diffusive behaviour in a fractal space. The city is self-organized to be more efficient in the space occupation, sharing more area of the city than would happen in a process governed by a standard Fokker-Planck dynamics, in which case the area increases linearly with the population size. 


The central result in the present work is Eq.~(\ref{eq:scalingexp}), which establishes the relations among the entropic index, the fractal dimension and the scaling exponent. All these quantities can be determined by independent methods so, at this point, a set of values for those parameters found in the Literature will be discussed. A more constrained analysis will be performed later on.

The value for the parameter $q$ appearing in the PPE can be calculated from analysis of information diffusion in urban areas and of epidemiologic analysis~\cite{Policarpo2023,Tirnakli2020}. The number of contacts per individual $n_c$ was associated with the entropic index by $n_c=(q-1)^{-1}$~\cite{Policarpo2023}. 
Studies about the number of contacts show that one can consider the range $3<n_c<10$ corresponding to the values $1/10<q-1<1/3$.~\cite{Policarpo2023,Tirnakli2020,Ghafoor2019}. Studies on the geographic shape of cities globally~\cite{Benguigui2000,Chen2017} provide city dimensions typically within the range of $1.4< d_f <1.9$, which corresponds, according to Eq.~(\ref{eq:chi-fractalgap}), to the range $1/20<q-1<3/10$. A systematic study of cities over all continents concluded that the scaling exponents lie within the range $2/3<\beta<1$~\cite{Burger2022}, which corresponds to the range $0<q-1<1/3$. Therefore, considering all the independent information on the range of the parameter $q$, one can estimate that $0<q-1<1/3$.


\begin{figure*}[t]
\centering  
\includegraphics[width=0.47\textwidth]{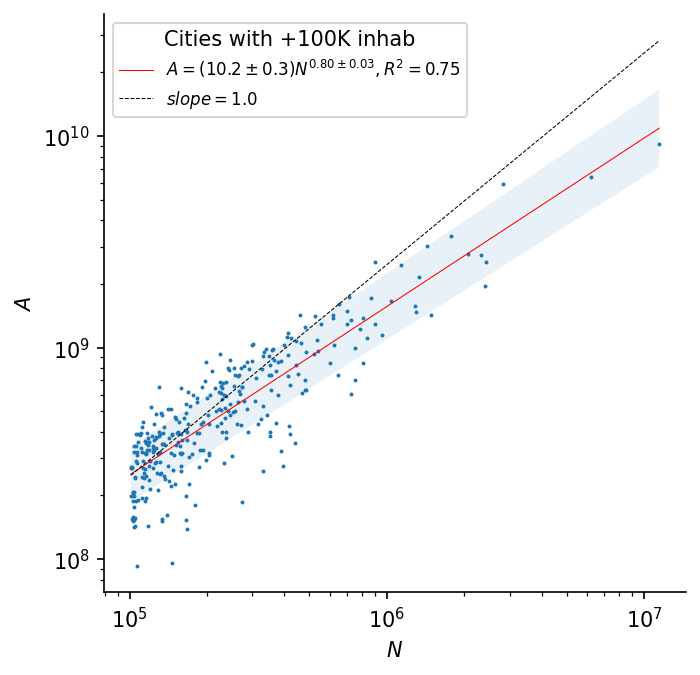}
\includegraphics[width=0.47\textwidth]{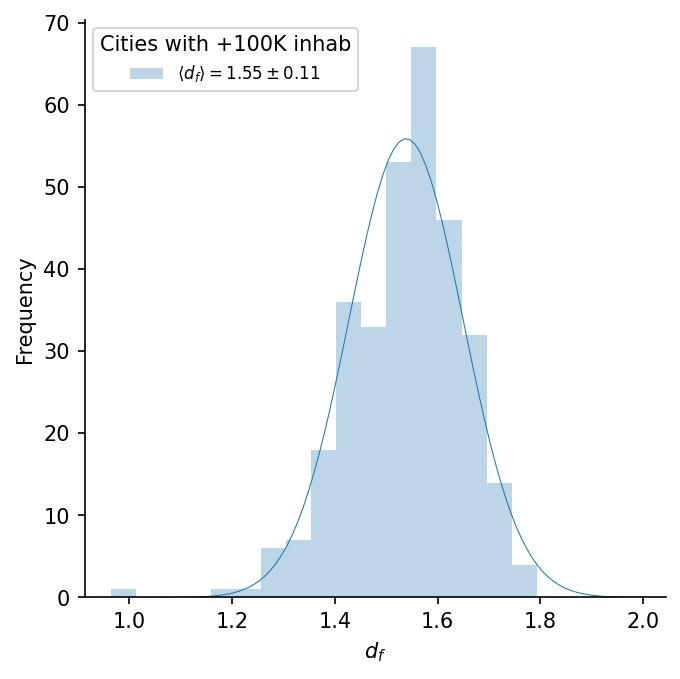}
 \caption{Study of the allometric relation for Brazilian cities with a population larger than $10^5$. The left panel shows a log-log plot of the urban area as a function of the population. The continuous (red) line represents the best fit to the data by using the allometric relation in Eq.~(\ref{eq:allometry}), resulting in $\beta=0.80 \pm 0.03$, and the black line indicates the linear behaviour ($\beta 
 = 1$), for the sake of comparison. 
  The right panel shows a histogram of the fractal dimension for Brazilian cities with population $N \ge 10^5$, resulting in an average $\langle d_f \rangle = 1.55 \pm 0.11$. A Gaussian curve was adjusted to the data for comparison. }
\label{fig:BetaDf}
\end{figure*}

So far, we have shown that considering the average values for the fractal dimension and the scaling exponent of the cities around the world, Eq.~(\ref{eq:scalingexp}) works properly. A more restrictive test is to observe the description of that formula to specific cities for which both the fractal and the scaling exponent are obtained.

\section{A case study: Brazilian cities}

As a case study, the verification of the theoretical relations derived here is done by studying the observed data of Brazilian cities. For a set of cities with a population larger than three hundred thousand, the allometric exponent $\beta$ was obtained by using information on the population size and the urban area. In addition, for the same cities, the fractal dimension of the urban area, $d_f$, is obtained by a process that considers the spatial distribution of crossing of roads in the city.

\begin{figure*}[t]
\centering   \includegraphics[width=0.46\textwidth]{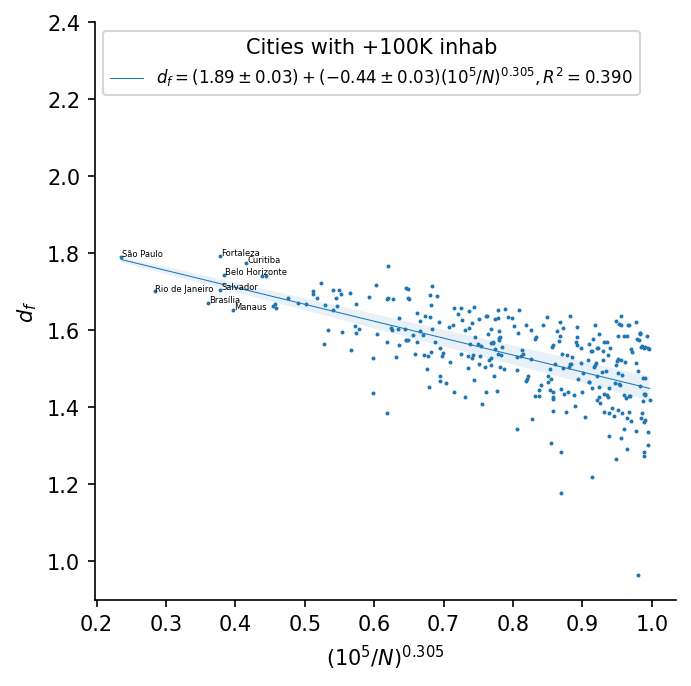}
\includegraphics[width=0.47\textwidth]{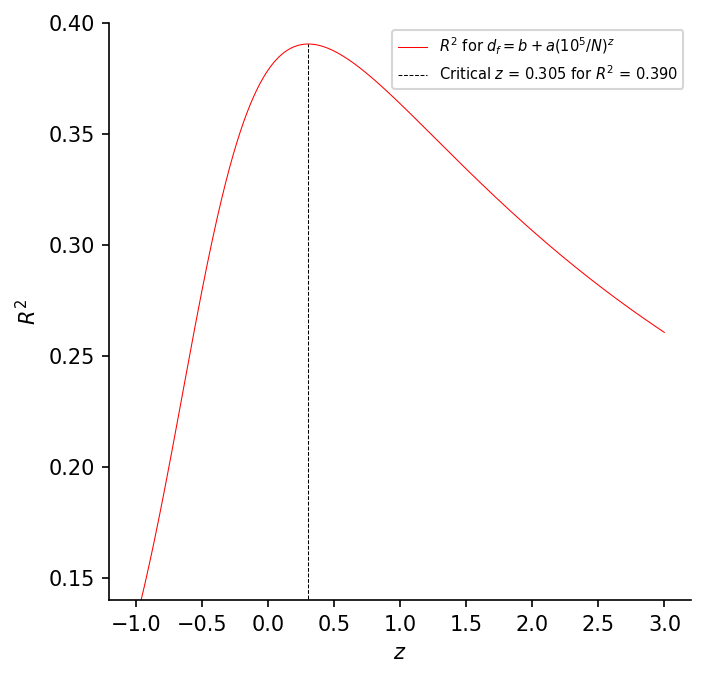}
 \caption{ Study of the fractal dimension for Brazilian cities with a population larger than $10^5$. The panel at the left represents the individual cities' fractal dimensions as a function of the rescaled inverse population size. The rescaling exponent was chosen to maximize the value for $R^2$ in the linear fit, as described in the right panel, where $R^2$ is plotted against the rescaling exponent $z$. 
 }
\label{fig:Df2beta}
\end{figure*}

The data used here for the evaluation of the allometric exponent is obtained from Refs.~\cite{IBGEarea2019, IBGEpopulation2022}. The data used to analyze the urban boundaries of the cities is taken from ~\cite{Embrapa2015}. Before calculating the fractal dimension, the data is processed by extracting the road intersections using the Python package ~\cite{OpenStreetMap}. For calculating the fractal dimensions, the box-counting method is applied to the road intersections by defining
a lattice. This method initially defines a $L \times L$ square with $L \sim \sqrt{A}$, where $A$ is the urban area of the city. 
The square is centred at the geometric centre of the road intersections of the city and encloses all the city's road intersections.

The fractal dimension is obtained by counting those boxes with side $\ell$ that contain at least one road intersection inside, $N_{\ell}$. The process is repeated iteratively for boxes with size
\begin{equation}
    \ell_i=\frac{L}{2^i} \,,
\end{equation}
with $i \in [2,8]$, a range that was found suitable for the set of cities considered here. The fractal dimension $d_f$ is obtained by using the formula
\begin{equation}
    d_f=\frac{\log(N_{\ell_i})}{\log(\ell_i)} \,.
\end{equation}
Typical plots of $N(\ell)~vs~\ell$ are shown in Fig.~\ref{fig:plotDf}. In the determination of $d_f$, only the linear region of each plot was considered, following the recommendations in Ref.~\cite{Molinero2021}.


As reported in the left panel of Fig.~\ref{fig:BetaDf}, the allometric relation analysis gives $\beta=0.80 \pm 0.03$, which lies within that range observed for cities around the world. The distribution of the fractal dimension is reported in the histogram shown in the right-hand panel of Fig.~\ref{fig:BetaDf}, where one can see an approximately normal distribution with the average at $\langle d_f \rangle=1.55 \pm 0.11$, which falls in the range observed for cities worldwide. Eq.~(\ref{eq:scalingexp}) represents the central result of the present work. It gives that $\langle d_f \rangle/(2 \beta)=0.97 \pm 0.07$, in agreement with the expected value and confirming that the theoretical approach can correctly predict the relation between the allometric exponent and the city's fractal dimension.

A more detailed analysis can be performed by considering the fractal dimension results for each of the cities in the present study. The left panel in Fig.~\ref{fig:Df2beta} displays the fractal dimension as a function of the rescaled inverse population size. The rescaling is done by using the critical exponent, $z=0.305$, that maximizes $R^2$ in the linear fitting as can be seen in the right panel of this figure. The continuous line is a linear fitting, which indicates a weak dependence of the fractal on the population size. Observe that for $N \rightarrow \infty$ (thermodynamical limit), the fractal dimension results to be $d_f=1.89 \pm 0.03$, showing that even for extremely large populations, the urban area remains a fractal structure. Interestingly, the asymptotic value of $d_f$ agrees with the fractal dimension of a percolation cluster in a two-dimensional space, which is $91/48$~\cite{Xu2014,Stauffer2003,Coniglio1989,Havlin}.

\begin{figure*}[t]
\centering   
\includegraphics[width=0.46\textwidth]{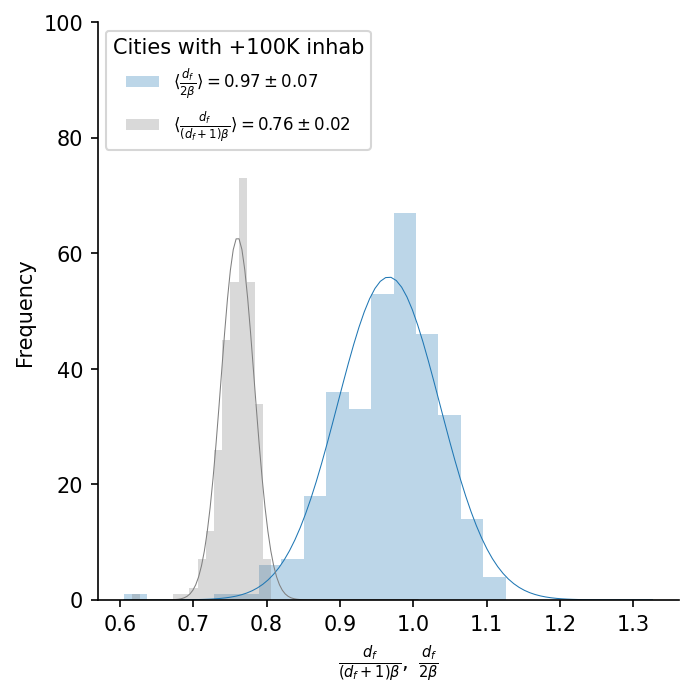}
\includegraphics[width=0.47\textwidth]{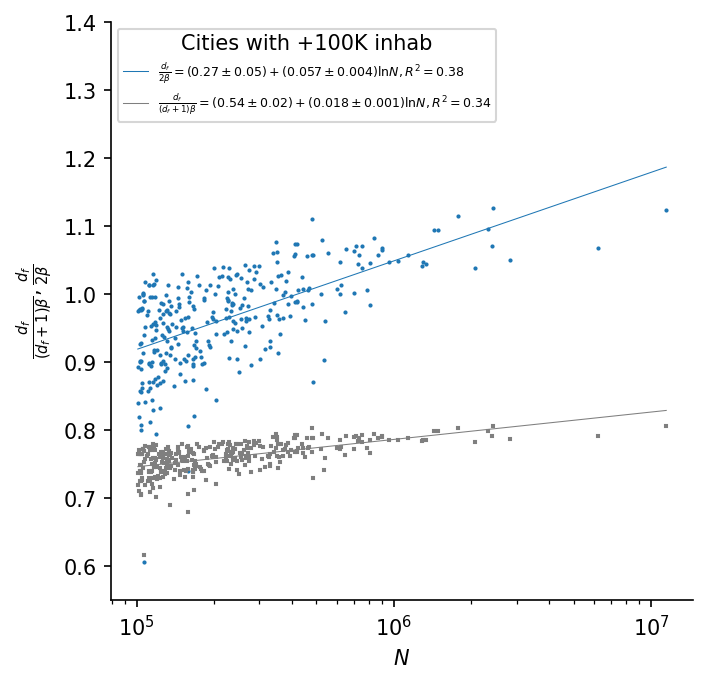} 
 \caption{Comparison between the results obtained with the fractal diffusion approach presented here (blue), and the result obtained by using Eq.~(\ref{eq:Bettencourt-Louf}) (black).  The left panel shows the histogram of the normalized relation between $\beta$ and $d_f$ by each model. For a successful prediction, the results should be around the unit. The right panel shows the behaviour of the ratios with the population size. A weak dependence on the population is observed.
  }
\label{fig:Df2BetaS}
\end{figure*}

The left panel in  Fig.~\ref{fig:Df2BetaS} displays the ratio $d_{f_i}/(2 \beta)$, where the fractal dimension of $i$th city is used instead of the average value. From the theory, it is expected that all data points spread around the unit. Observe that the weak dependence of the ratio $d_{f_i}/(2 \beta)$ with the population $N$ is similar to that of $d_f$ with the population size. It corroborates that the fractal dimension is not a truly universal aspect of the cities but shows a weak dependence on the population size~\cite{Batty2013}. According to Eq.~(\ref{eq:scalingexp}), also $\beta$ should present a dependence on the city's population, but in the present analysis, a constant value was adopted. Such weak dependence of $d_f$ could be the cause of the slightly asymmetric distribution in Fig.~\ref{fig:BetaDf}, but the present analysis is not intended to give a definite answer in this regard. 

It must be emphasized that the present approach to the cities' allometric relation is rather different from the existing models, although leading to similar results. It is completely based on the diffusion process in fractal spaces, avoiding assumptions about individual behaviour. On the contrary, Bettencourt's~\cite{Bettencourt2013}
and Ribeiro et al.  \cite{Ribeiro2024} models are strongly based on the assumption of efficiency determining the approximate allometric exponent but need the introduction of a fractal dimension. The Loaf-Barthelemmy model~\cite{Louf2014} is based on traffic and commutation time.

Bettencourt's, Ribeiro et al.'s,   and Loaf-Barthelemmy's models give the same formula relating the allometric exponent and the fractal dimension, namely,
\begin{equation}
    \beta=\frac{d_f}{d_f+1}\,. \label{eq:Bettencourt-Louf}
\end{equation}
Comparing with Eq.~(\ref{eq:scalingexp}) for $d=2$, giving $\beta=d_f/2$, and considering the expected value for the fractal dimension, $1.4 < d_f < 1.9$, it results that the present model will always give an exponent larger than that given by the other models discussed here. 

In the left panel of Fig.~\ref{fig:Df2BetaS}, the accuracy of the prediction given by Eq.~(\ref{eq:Bettencourt-Louf}) is compared with that for the present approach, given by Eq.~(\ref{eq:scalingexp}). It is observed a significant deviation from the expected unit value for the case of Eq.~(\ref{eq:Bettencourt-Louf}), showing that the fractal approach is more accurate for the cities used in the present study. The right panel in Fig;~\ref{fig:Df2BetaS} displays the behaviour of the relevant ratios for the different models as a function of the population size.

\section{City's attraction potential}

There is another aspect of the dynamics approach that deserves further consideration. The relation between the occupied area and the population size involves, beyond the fractal aspects of the city, socioeconomic factors that depend on the attractiveness of the city, which motivates the population to move near to its centre. In the dynamical approach employed in the present work, this is controlled by the potential associated with the parameter $\gamma_2$. Hence, this work considers the general case of a power-law attractive force of the form $\gamma_2 r^{\omega}$ replacing the harmonic potential. The  stationary distribution has a $q$-exponential form, being given by
 \begin{equation}
      f_q(\mathbf{r})= \frac{1}{N(t)} e_q\left[- \frac{1}{2} \left( \frac{r}{\sigma_{\infty,\omega}} \right)^{2\alpha}\right]\,, \label{eq:qNgauss-m}
 \end{equation}
 where
 \begin{equation}
     \alpha = \frac{1 + \omega}{2} \,.
 \end{equation}
The stationary width is given by
\begin{equation}
    \sigma_{\infty,\omega} \equiv \ell_{q,\omega} \, \kappa^{1/(\alpha\nu)} \,, \quad
    \kappa \equiv \alpha (2-q) (2\pi \chi_{q,\omega})^{\frac{d}{2}(q-1)} \frac{B}{\gamma_2 \ell_{q,\omega}^{2\alpha}}  \,, \label{eq:sigma-kappa-beta-m} 
\end{equation}
with
\begin{equation}
    \chi_{q,\omega} \equiv  \frac{1}{2} \left( \frac{2}{q-1} \right)^{\frac{1}{\alpha}} \left(  \frac{ \Gamma\left( 1 + \frac{d}{2\alpha} \right)  }{ \Gamma\left( 1 + \frac{d}{2}\right)  } \frac{ \Gamma\left(  \frac{1}{q-1} - \frac{d}{2\alpha} \right)  }{ \Gamma\left( \frac{1}{q-1} \right)  } \right)^{\frac{2}{d}}  \,,
\end{equation}
and 
\begin{equation}
    \nu \equiv  2  - \frac{d}{\alpha} (q-1) \,. \label{eq:nu}
\end{equation}
The population size results in
\begin{equation}
 N_{\infty,\omega} = \left( \sqrt{2\pi \chi_{q,\omega}} \frac{\sigma_{\infty,\omega}}{\ell_{q,\omega}} \right)^{d} \,. \label{eq:Nsigmat-beta-m} 
\end{equation}

Following the same reasoning used above for the case $\omega = 1$, it is found that the population size will be independent of $q$ if
\begin{equation}
    \ell_{q,\omega} \propto \ell_{\BG}^{\nu/2} \,,
\end{equation}
where $\ell_{\BG} \equiv \ell_{q=1,\omega=1}$ corresponds to the case where the city area is proportional to the population size. This scaling property leads to $A_{q,\omega} = \ell_{q,\omega}^d =A_{\BG}^{\nu/2}$, therefore 
\begin{equation}
   A_{q,\omega} \propto N^{\nu/2} \,.\label{eq:qbetaarea}
\end{equation}
For a given scaling exponent and fractal dimension, the power-law exponent $\alpha$ can be easily obtained from Eq.~(\ref{eq:qbetaarea}), resulting
\begin{equation}
    \alpha =\frac{d-d_f}{2 (1 -\beta)} \,.  \label{eq:alpha}
\end{equation}
The parameter $\alpha$ can be useful to understand the few cases where $\beta \lesssim 2/3$, corresponding to $q-1 \gtrsim 1/3$.

\begin{figure*}[t]
\centering   
\includegraphics[width=0.46\textwidth]{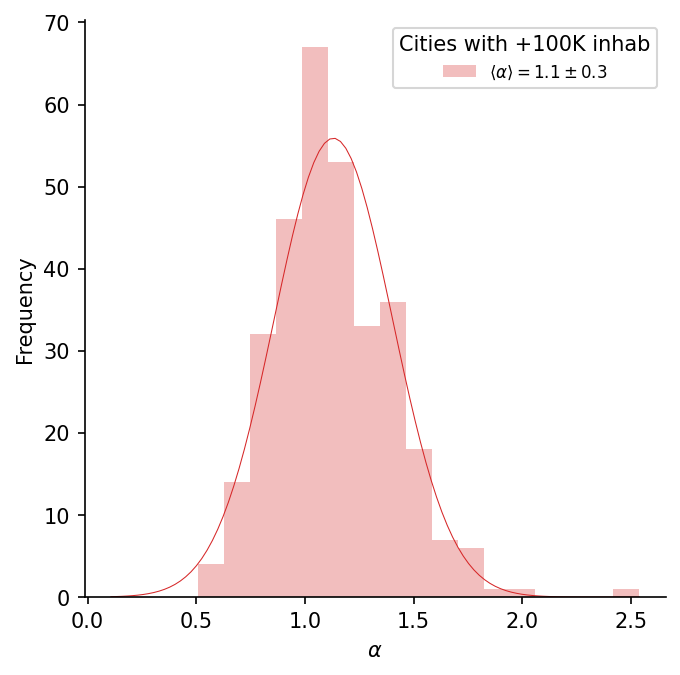}
\includegraphics[width=0.46\textwidth]{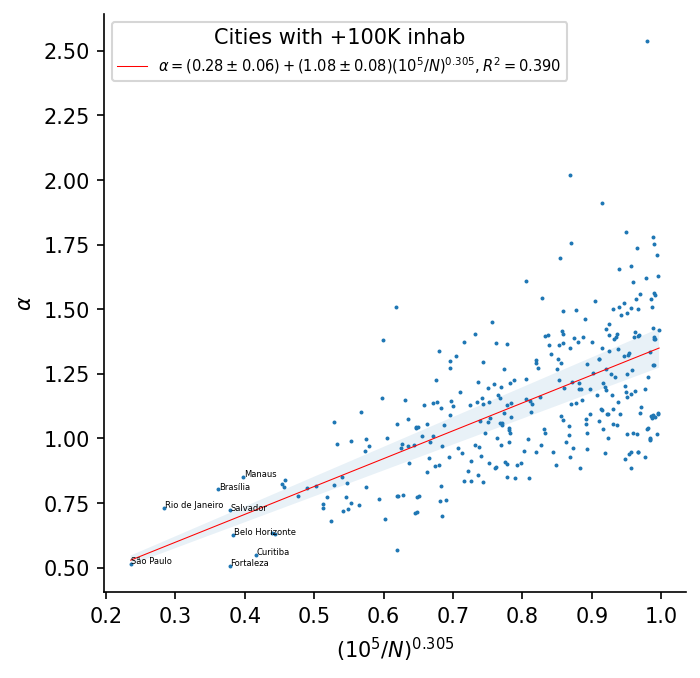}
 \caption{Results for $\alpha$, as calculated by Eq.~(\ref{eq:alpha}). The histogram of the individual city's results is displayed in the left panel. The right panel shows the values of $\alpha$ as a function of the rescaled population size.
  }
\label{fig:alpha}
\end{figure*}

The value for $\alpha$ was verified for the set of Brazilian cities used in the case study presented here. The result, as observed in the left panel in Fig.~\ref{fig:alpha} shows that $\langle \alpha \rangle = 1.1 \pm 0.3$ is a value that agrees with the harmonic potential adopted in the previous sections. However, for a constant value $\beta$, the parameter $\alpha$ varies with the population size. In the right panel of Fig.~\ref{fig:alpha}, the value of the parameter $\alpha$ is plotted against the rescaled inverse population size. The rescaling exponent is obtained in the same way as indicated previously, and because $\alpha$ has a linear dependence on $d_f$, the critical exponent for the rescaling is the same, $z=0.305$. Observe that for $N \rightarrow \infty$, $\alpha \rightarrow 0.28 \pm 0.06$, indicating that even in the thermodynamical limit the city's attractive potential remains a power-law function of the distance to the centre of the urban area.

The plots for this study are presented in the Supplementary Material.

%
%
%

\section{Conclusions}

In summary, the dynamical approach used to characterize urban scaling underscores the significance of the Plastino-Plasti\-no Equation in addressing the anomalous diffusion linked to the intricacies of urban life and the fractal geometry of cities. The connections between this nonlinear equation, fractal and fractional calculus, and nonextensive statistics yield Eq.~(\ref{eq:chi-fractalgap}). This equation establishes a relationship between the fractal dimension of the city, the entropic index, and the scaling exponent. While the fractal dimension elucidates the complex geometry of the urban space~\cite{MGD2024}, the entropic index can be linked to the social activities of individuals~\cite{DeppmanPLOSOne2021}, providing a comprehensive understanding of the correlations between infrastructure and socioeconomic behaviour that shape urban life, aligning with previous works~\cite{Xu2021-ah,Grabowicz2014}.

The dynamics in fractal spaces provide only two parameters for describing the system behaviour, namely, the entropic index and the attractive potential exponent. 
The latter represents the attractiveness of the city. The interplay between these two parameters can precisely account for the scaling exponent observed in any city, highlighting the intricate relationships between infrastructure-related issues and social interactions. The results indicate that a weak dependence of both fractal dimension and allometric exponents with the population size is necessary. Overall, the dynamical approach gives fair prediction to the population stationary distribution of cities. 

For the set of Brazilian cities used in the study of the case presented in this work as a test for the prediction of the fractal diffusion approach, the results obtained are better than the best-known models in the Literature~\cite{Bettencourt2013,Louf2014}. However, in all cases, the fundamental origin of the fractal space remains obscure but the present theoretical approach evidences that the relation between the fractal dimension and the allometric exponent is associated with basic geometric and diffusion aspects. The present model also allows for studying the temporal evolution of the population distribution in the urban area, offering new methods for testing the predictions given by the theoretical approach.

This work opens up the possibility of addressing the dynamical aspects of cities and offers new perspectives for understanding the origins of fractality in urban life. Systematic analyses of the two parameters provided by the theory can elucidate the complex connections between social behaviour and the city's design. The methods presented here can be of help in the design of growing infrastructure in cities~\cite{He2023}, and the promotion of economic growth~\cite{Zhang2023,Chen2023}. Future research may examine the relationship between the number of contacts and the shared area of the city associated with the scaling exponent. Investigating the determining features of the human mind that underlie the emergence of fractal behaviour remains an intriguing area for scientific development.  The {\it Science of Cities} is necessarily interdisciplinary, encompassing physical, mathematical, sociological and philosophical aspects. In the latter areas, progress has been made in understanding the implications of a complex approach to social behaviour in modern society~\cite{Morin2014-zr,Donati2023}.

\section{Acknowledgements}

A.D. is supported by Conselho Nacional de Desenvolvimento
Cient\'{\i}fico e Tecnol\'ogico (CNPq-Brazil), grant 306093/\\2022-7. The work of E.M. is supported by the project PID2020-114767GB-I00 and by the Ram\'on y Cajal Program under Grant RYC-2016-20678 funded by MCIN/AEI/10.13039/501100011\-033 and by ``FSE Investing in your future'', by Junta de Andaluc\'{\i}a under Grant FQM-225, and by the ``Pr\'orrogas de Contratos Ram\'on y Cajal'' Program of the University of Granada. C.T. is partially supported by the Brazilian Agencies CNPq and Faperj. 
FLR thanks CNPq (grant numbers 403139/2021-0 and 424686/2021-0) and Fapemig (grant number APQ-00829-21) for financial support.

\bibliographystyle{ieeetr}
\bibliography{DynamicCity}

\pagebreak[4]

\section{Supplementary Material (will not appear in the main text}

\begin{figure*}[t]
\centering  
\includegraphics[width=0.45\textwidth]{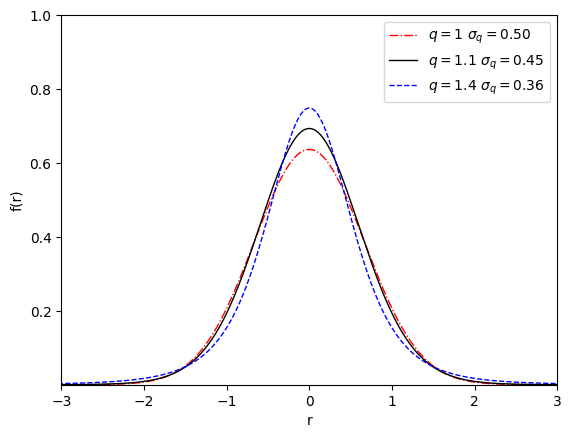}  
\hspace{1cm}  
\includegraphics[width=0.45\textwidth]{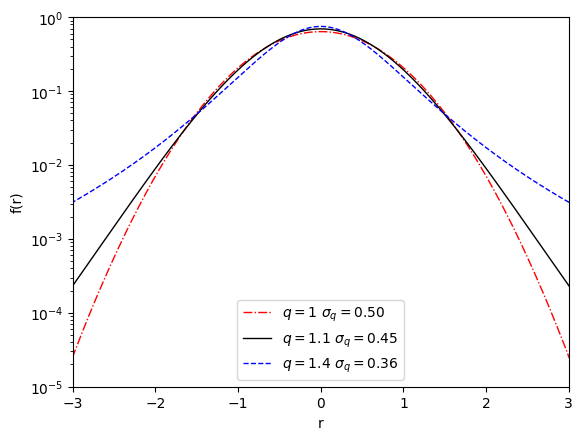}  
 \caption{Plots of the population distribution function profiles for different values of $q=1,~1.1,~1.4$ in the PPE. They correspond, respectively, to the fractal gap $d_f=2,~1.45,~1.30$. The horizontal axis gives the distance $r$ to the city centre in arbitrary units, and the vertical axis gives the distribution function, in arbitrary units. The left panel shows a linear-linear plot, while the right plot shows a log-linear plot. 
 }
\label{fig:qgauss}
\end{figure*}

The correlations in the city's structure imply the use of PPE instead of the FPE. The solutions for the former equation are $q$-Gaussians, in contrast with the solutions for the latter one, which are Gaussian distributions.
Fig.~\ref{fig:qgauss} shows the $q$-Gaussian distributions for different values of $q$ and width $\sigma$ in the stationary regime that is obtained by using Eq.~(\ref{eq:scaling-sigmaS}) for a fixed population size, $N$. The case $q=1$ corresponds to the standard FPE solution, which is a Gaussian distribution, and it is compared with the PPE solutions for two different values of $q$. The plots show that in the central region around $r=0$ the $q$-Gaussian
present a more pronounced peak, which results from the narrower distribution width due to fractal effects in the city dynamics. However, the $q$-Gaussians are fat-tailed, resulting in larger populations at the borders of the city.
It means that the area occupied by the population in the city is smaller than it would be the case if the population was randomly distributed around the centre. In terms of the entropic index, the result evidences the non-additivity of the city configurations. The value $q \ne 1$ explains why, upon adding a group of individuals to an existing city, the area of the new city will not be the simple sum of the area initially occupied by each group.

\begin{figure*}[t]
\centering  
\includegraphics[width=1\textwidth]{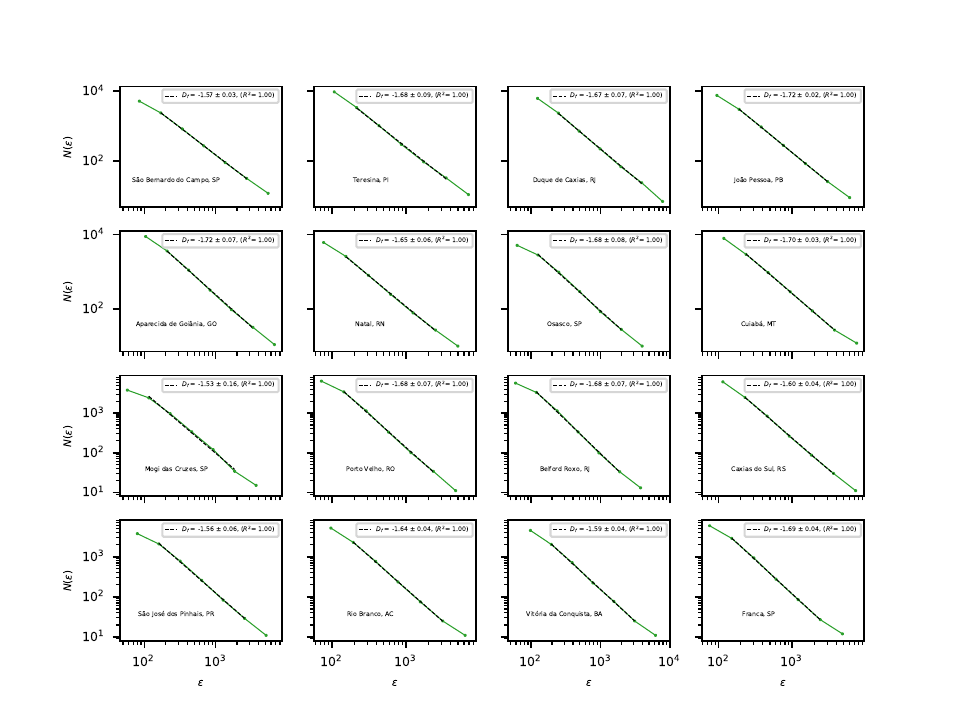}  
\hspace{1cm}  
 \caption{Sample of the fittings for the determination of the cities' fractal dimension.  The traced line indicates the best linear fit and the region considered for the fittings.}
\label{fig:BetaS}
\end{figure*}

Fig.~\ref{fig:BetaS} shows a sample of the 319 plots to obtain the Brazilian cities' dimensions that were used for the study case.

\end{document}